\documentstyle[preprint,aps,graphicx]{revtex}
\tightenlines
\begin{document}

\title{Phenomenological Equations of State for the Quark-Gluon Plasma}

\author{Peter N. Meisinger, Travis R. Miller and Michael C. Ogilvie}

\address{ Dept. of Physics
Washington University
St. Louis, MO 63130} 

\date{\today}

\maketitle

\begin{abstract}
Two phenomenological models describing an $SU(N)$ quark-gluon
plasma are presented. The first is obtained from high temperature expansions
of the free energy of a massive gluon, while the second is derived by
demanding color neutrality over a certain length scale. Each model has a
single free parameter, exhibits behavior similar to lattice simulations over
the range $T_{d}-5T_{d}$, and has the correct blackbody behavior for large
temperatures. The $N=2$ deconfinement transition is second order in both
models, while $N=3$,$4$, and $5$ are first order. Both models
appear to have a smooth large-$N\,\ $limit.
For $N \geq 4$, it is shown that the trace of the Polyakov loop
is insufficient to characterize the phase structure; the free energy is best
described using the eigenvalues of the Polyakov loop.
In both models, the confined
phase is characterized by a mutual repulsion of Polyakov loop
eigenvalues that makes the Polyakov loop expectation value zero. In the
deconfined phase, the rotation of the eigenvalues in the complex plane
towards $1$ is responsible for the approach to the blackbody limit over the
range $T_{d}-5T_{d}$. 
The addition of massless quarks in $SU(3)$ breaks $%
Z(3)\,$symmetry weakly and eliminates the 
deconfining phase transition. In contrast, a
first-order phase transition persists with sufficiently heavy quarks.

\end{abstract}


\section{Introduction}
\label{sec:intro}

\vspace{1pt}The equation of state for the quark-gluon plasma is of great
interest in several different areas of physics. In heavy ion physics,
astrophysics, and cosmology, the equation of state is needed as input. There
are two first-principles methods to obtain the equation of state:
perturbation theory and lattice simulations.
Only lattice gauge theory techniques are able to
determine the equation of
state directly over all temperatures
but researchers in other fields have typically contented
themselves with the extraction of a few important parameters from lattice
results. When the equation of state is needed, the bag model equation of
state \cite{Cleymans:1986wb}
is very often used, even though it is a poor representation of lattice
results.

\vspace{1pt}As better alternatives, we have developed two simple models for
the $SU(N)$ gluon plasma equation of state. Both models are obtained by
combining simple phenomenological ideas with $Z(N)$\ symmetry and the known
features of the perturbative equation of state. These models have many
desirable properties, and exhibit thermodynamic behavior similar to that
obtained from lattice simulations. In particular, they give a reasonable
picture of the crucial region from the deconfinement temperature 
$T_{d}$ to $5T_{d}$ which has not been
obtained by other means. The phase transition in these models is second
order for $N=2$, and first order for $N=3$, $4$, and $5$. It appears that
the large-$N$ limit will be smooth in both models. Quark effects are easily
included.

\vspace{1pt}In both models, the eigenvalues of the Polyakov loop 
determine all thermodynamic properties.
The
Polyakov loop is defined for Euclidean finite temperature gauge theories as 
\begin{equation}
P(\overrightarrow{x})={\cal T}{\Large \,\exp }\left[
ig\int_{0}^{1/T}d\tau \,A_{0}(\overrightarrow{x},\tau )\right] 
\end{equation}
where $T\,$is the temperature, and ${\cal T}$ denotes Euclidean time
ordering. The Polyakov loop is is the natural order parameter of the
deconfinement transition in pure gauge theories. Its trace in the
fundamental representation can be related to the free energy $F$\ of a heavy
source: 
\begin{equation}
e^{-F/T}=\left\langle Tr_{F}\,P(\overrightarrow{x})\right\rangle 
\end{equation}
where the expectation value is taken over a thermal ensemble of states.
It has been known for some time that
$\left\langle Tr_{F}\,P(\overrightarrow{x})\right\rangle$
is of fundamental importance in describing the
deconfining phase transition \cite{Yaffe:1982qf}.
In a pure gauge theory, the action is invariant under a global symmetry
associated with the center $C$ of the gauge group, but the Polyakov loop is
not invariant. If $z\in C$, then the symmetry requires $z\left\langle
Tr_{F}\,P(\overrightarrow{x})\right\rangle =\left\langle Tr_{F}\,P(%
\overrightarrow{x})\right\rangle $, which implies $\left\langle Tr_{F}\,P(%
\overrightarrow{x})\right\rangle =0$. This is in turn interpreted as $%
F=\infty $, indicating confinement. 
In the phenomenological models developed here, the eigenvalues of the
Polyakov loop are the essential degrees of freedom rather than
$\left\langle Tr_{F}\,P(\overrightarrow{x})\right\rangle$ alone,
a possibility also recently explored by Pisarski
\cite{Pisarski:2000eq}.
The Polyakov loop $P$ is unitary for $%
SU(N)$ gauge theories; we will denote its eigenvalues in the fundmental
representation by 
\begin{equation}
P_{jk}=\exp \left( i\theta _{j}\,\right) \delta _{jk} 
\end{equation}
after a diagonalizing unitary transformation. The eigenvalues in other
representations will be linear combinations of the phase factors
$\exp (i\theta_j)$. For simplicity, we will refer to both $\exp (i\theta_j)$
and $\theta_j$ as eigenvalues, relying on context to differentiate them.

In our models, 
we make a mean-field assumption that the Polyakov loop eigenvalues
are constant throughout space, and assume that 
the free energy $f$ is a function of the eigenvalues.
Confinement is obtained from a set of eigenvalues which
make $\left\langle Tr_{F}\,P(\overrightarrow{x})\right\rangle =0$. As we
show below, this is naturally obtained by a uniform distribution of
eigenvalues around the unit circle, constrained by the unitary of $P$. In
other words, confinement at low temperatures is a consequence of eigenvalue
repulsion. In a pure gauge theory below the deconfinement temperature $T_{d}$%
, the eigenvalues are frozen in this uniform distribution. 
As $T$ moves upward from $T_d$,
the eigenvalues of the Polyakov loop rotate towards $\theta =0$ or
one of its $Z(N)\,$equivalents, and $P$ moves towards an element of the
center. In the case of a first order transition, the eigenvalues jump at $%
T_{d}$. In the models developed below,
it is this motion of the eigenvalues which is
responsible for the approach to the blackbody limit over the range $%
T_{d}-5T_{d}$.

The next section briefly derives the conventional one-loop expression for
the free energy of gluons in a constant Polyakov loop background. Using this
expression, section~\ref{sec:landau}
explains in detail why a naive Landau-Ginsberg
treatment of deconfinement based on $Tr_{F}\,P(\overrightarrow{x})$ as a
single complex order parameter fails for $SU(4)$ and higher. 
Sections~\ref{sec:modela} and
~\ref{sec:modelb} introduce model A and model B, respectively.
Model A is particularly
tractable analytically in the cases of $SU(2)$ and $SU(3)$. 
Section~\ref{sec:thermo}
presents results for the pressure $p$, the internal energy $\varepsilon $,
and the interaction measure $\Delta $ for both models for $N=2$, $3$, $4$,
and $5$.
Section~\ref{sec:quarks} considers the effects
of massive and light quarks in model A for $SU(3)$. A final section
discusses our results.

\section{Perturbative EoS at high T}
\label{sec:perturb}

There are several reasons for beginning with the perturbative expression for
the free energy. As a consequence of asymptotic freedom, the perturbative
expression for the free energy as a function of the Polyakov loop
eigenvalues will be valid at sufficiently high temperatures.
Any purely perturbative
calculation will give a free energy of the form 
\begin{equation}
f_{pert}=T^{4}F(P,g(T)) 
\end{equation}
where $F$ depends directly only on dimensionless variables: the Polyakov
loop $P$ and the running coupling constant $g(T)$. The leading order
behavior, which is sufficient for our needs, is independent of $g(T)$ and
gives the blackbody behavior expected at high temperature. In addition, the
perturbative expression for the free energy will give us significant
insights into the structure and interpretation of possible terms in the free
energy.

\vspace{1pt}At one loop, the free energy for gluons in a constant $A_{0}$
background can be written as
\begin{equation}
f_{pert}(\theta )=\ln \left[ \det \left( -D_{adj}^{2}\right) \right] 
\end{equation}
where $D_{adj}$ is the covariant derivative acting on fields in the adjoint
representation. This can be written in an unregularized form as 
\begin{equation}
f_{pert}(\theta )=\sum_{n\in Z}\int \frac{d^{3}k}{(2\pi )^{3}}tr\,_{A}\,\ln
\left[ \left( \frac{2\pi n}{\beta }-A_{0}\right) ^{2}+\overrightarrow{k}%
^{2}\right] 
\end{equation}
where the trace is in the adjoint representation
and $\beta = 1/T$.
Using a standard
identity \cite{GandR},
this can be written in a familiar form as the sum of a
zero-temperature contribution, which is divergent, and a finite temperature
contribution: 
\begin{equation}
f_{pert}(\theta )=2\sum_{n\in Z}\int \frac{d^{3}k}{(2\pi )^{3}}%
tr_{A}\,\,\left\{ \frac{1}{2}\omega _{k}+\ln \left[ 1-\exp \left( -\beta
\omega _{k}-i\beta A_{0}\right) \right] \right\} 
\end{equation}
where $\omega _{k}\equiv \left| \overrightarrow{k}\right| $ and the factor
of $2$ comes from the sum over polarization states. Disregarding the zero
temperature contribution, the free energy density can be written as 
\begin{equation}
f_{pert}\left( \theta \right) =\frac{1}{\beta }\sum_{j,k=1}^{N}2(1-\frac{1}{N%
}\delta _{jk})\int \frac{d^{3}k}{(2\pi )^{3}}\ln \left[ 1-e^{-\beta
\omega_{k}+i\Delta \theta _{jk}}\right] 
\end{equation}
where the factor involving the Kronecker $\delta $ projects out the singlet
state. We have defined the differences of the fundamental representation
eigenvalues as $\Delta \theta _{jk}$ $\equiv \theta _{j}-\theta _{k}$; they
are just the $N^{2}-1$ eigenvalues in the adjoint representation. It is
instructive to expand the logarithm: 
\begin{equation}
f_{pert}\left( \theta \right) =-\frac{1}{\beta }\sum_{j,k=1}^{N}2(1-\frac{1}{%
N}\delta _{jk})\int \frac{d^{3}k}{(2\pi )^{3}}\sum_{n=1}^{\infty }\frac{1}{n}%
e^{-n\beta \omega_{k}+in\Delta \theta _{jk}}\text{.}
\end{equation}
Each factor of $n$ is associated with paths that wrap around space-time $n\,$%
times in the Euclidean time direction \cite{Meisinger:1995qr}.
The integral over $k$
can be performed, yielding 
\begin{equation}
f_{pert}\left( \theta \right) =-\frac{2}{\pi ^{2}\beta ^{4}}%
\sum_{j,k=1}^{N}(1-\frac{1}{N}\delta _{jk})\sum_{n=1}^{\infty }\frac{1}{n^{4}%
}e^{in\Delta \theta _{jk}}
\end{equation}
Note that $f_{g}$ is real because $\Delta \theta _{jk}=-\Delta \theta _{kj}$%
. The summation over $n$ can be done exactly, giving 
\begin{equation}
f_{pert}\left( \theta \right) =\frac{2T^{4}}{\pi ^{2}}\sum_{j,k=1}^{N}(1-%
\frac{1}{N}\delta _{jk})\left[ \frac{\pi ^{4}}{3}B_{4}\left( \left| \Delta
\theta _{jk}\right| _{2\pi }/2\pi \right) \right] 
\end{equation}
where $\left| \Delta \theta \right| _{2\pi }\equiv \left| \Delta \theta
\right| \,mod\,2\pi $ and $B_{4}(x)=x^{4}-2x^{3}+x^{2}-\frac{1}{30}\,$%
is the fourth Bernoulli polynomial. This leads immediately to
\cite{Gross:1981br,Weiss:1981rj,Weiss:1982ev}
\begin{equation}
f_{pert}(\theta )=-2T^{4}\sum_{j,k=1}^{N}(1-\frac{1}{N}\delta _{jk})\left[ 
\frac{\pi ^{2}}{90}-\frac{1}{48\pi ^{2}}\left| \Delta \theta _{jk}\right|
_{2\pi }^{2}\left( \left| \Delta \theta _{jk}\right| _{2\pi }-2\pi \right)
^{2}\right] 
\end{equation}
which reduces to
the usual black body formula in the case $A_{0}=0$. The free
energy is a smooth polynomial in the $\Delta \theta $ variables, except at
the points $\Delta \theta =0,\pm 2\pi ,\pm 4\pi ,...$ where the $mod$
function acts to maintain the periodicity which is manifest in the original
form of $f_{pert}$.

The minimum of $f_{pert}$ occurs at $A_{0}=0$ and values related to $A_{0}=0$
by $Z(N)$ symmetries for all values of the temperature $T$. We can
characterize these solutions as 
\begin{equation}
\theta _{j}=\frac{2\pi n}{N}\text{ \ \ \ }n=0,..,\left( N-1\right) \text{.} 
\end{equation}
The one-loop expression for the free energy of gluons propagating in the
background of a non-trivial, but constant, Polyakov loop thus predicts a gas
of gluons would always be in the deconfined phase. There is no indication
that higher orders in perturbation theory modify this result
\cite{Gocksch:1993iy}.

\section{Failings of the Landau Approach}
\label{sec:landau}

We define $P_{F}$ to be the expectation value of the trace of
the Polyakov loop in the fundamental representation,
$\left\langle Tr_{F}\,P(\overrightarrow{x})\right\rangle$.
Because $P_{F}$ 
is the order parameter for the deconfining phase transition
in pure gauge theories, it is natural to attempt to construct a Landau
theory for the deconfining phase transition as a polynomial in $P_{F}$
\cite{Svetitsky:1986ye,Meyer-Ortmanns:1996ea}.

An obvious
form for the Landau free energy $f_{L}$ is 
\begin{equation}
f_{L}=a_{2}P_{F}^{\ast }P_{F}+\frac{a_{4}}{2}\left( P_{F}^{\ast
}P_{F}\right) ^{2}+\frac{a_{6}}{3}\left( P_{F}^{\ast }P_{F}\right) ^{3}+%
\frac{b_{N}}{N}(P_{F}^{N}+P_{F}^{\ast N})
\end{equation}
where the coefficients are all real and temperature-dependent. Most of the
terms in $f_{L}$ are invariant under a $U(1)$ symmetry $P_{F}\rightarrow
e^{i\alpha }P_{F}$. The term $b_{N}(P_{F}^{N}+P_{F}^{\ast N})$ breaks this
symmetry down to $Z(N)$, the correct symmetry group for the $SU(N)$
deconfining transition. An exception occurs for $SU(2)$: $P_{F}$ is real, so
the $b_{2}$ term can be absorbed into $a_{2}$, and $f_{L}$ has only a $Z(2)$
invariance. While additional terms can be added to $f_{L}$, this is the
minimal form necessary to reproduce all known behavior in $3+1$ dimensions.
Models of the lattice $SU(N)$ deconfinement transition
valid in the strong-coupling limit indicate a first-order transition
for all $N \geq 3$ \cite{Green:1984sd,Ogilvie:1984ss}; there is
also an argument based on Schwinger-Dyson equations that the
$N=\infty$ limit also has a first-order transition
\cite{Gocksch:1983en}.
Recent lattice simulations of 
$SU(4)$ at finite temperature provide very good evidence that 
the $SU(4)$ transitition is first order \cite{Wingate:2001bb}.
The deconfining $SU(N)$ gauge transition can be first order for all $N \geq 3$
with the simple Landau free energy $f_{L}$  only if
the coefficients are carefully
chosen functions of the temperature. In the large $N$ limit, these
coefficients must scale in such a way that $f_{L}\propto N^{2}$.

There are several problems associated with using $f_{L}$ to obtain an
equation of state over a wide range of temperatures.
First, there are generally at least four unknown
coefficients to be determined. Because $P_{F}$ is dimensionless, the
coefficients are all of dimension four in $3+1$ dimensions. We can safely
assume that they grow no quicker than $T^{4}$ for large $T$, in accord with
the usual black body behavior. The standard Landau assumption would be that
each coefficient is a polynomial in $T$ of degree four or less. This gives
us generically $20$ undetermined parameters. Without additional assumptions,
it is impossible to make useful progress.

Secondly, this form of the free energy includes only part of the known
high-temperature physics. As we have seen in the previous section, the
high-temperature partition function depends on terms of the form $%
Tr_{A}\left( P^{n}\right) $ where $n$ can be associated with a gluon
trajectory winding around space-time $n$ times in the temporal direction. 
Because $%
Tr_{F}\left( P\right) Tr_{F}\left( P\right) ^{\ast }=Tr_{A}\left( P\right)
+1 $, the $a_{2}$ term is associated with the $n=1$ gluon trajectories, and
so on. We see that a polynomial in $P_{F}$ does not take into account all of
the high-temperature perturbative result.

Finally, the most general form of the free energy does not depend on $P\,$%
solely through $P_F$ and $P_F^{*}$. The high-temperature
perturbative free energy $f_{pert}$ illustrates this point. It is not a
function solely of $P_{F}$ and its conjugate, and cannot be written as an
infinite series in $P_{F}$ and $P_{F}^{\ast }$. A simple example will
illustrate this point. Consider two diagonal matrices lying in $SU(4)$,
defined by 
\begin{equation}
U_{1}=\left( 
\begin{array}{llll}
e^{i\pi /2} &  &  &  \\ 
& e^{i\pi /2} &  &  \\ 
&  & e^{-i\pi /2} &  \\ 
&  &  & e^{-i\pi /2}
\end{array}
\right) 
\end{equation}

and\vspace{1pt} 
\begin{equation}
U_{2}=\left( 
\begin{array}{llll}
e^{i\pi /4} &  &  &  \\ 
& e^{i3\pi /4} &  &  \\ 
&  & e^{i5\pi /4} &  \\ 
&  &  & e^{i7\pi /4}
\end{array}
\right) 
\end{equation}
Both $U_{1}$ and $U_{2}$ have zero trace in the fundamental representation,
yet the traces of their square are differenct: $Tr_{F}\,U_{1}^{2}=-4$ and $%
Tr_{F}\,U_{2}^{2}=0$, establishing that $Tr_{F}\,U^{2}$ cannot be a function
solely of $Tr_{F}\,U$. Explicit calculation shows furthermore that $%
f_{pert}\left( U_{1}\right) \neq f_{pert}\left( U_{2}\right) $.

At first sight, this may seem to contradict two standard results: a) the
characters form a complete, in fact orthogonal, basis on class functions; b)
all characters may be obtained from the fundamental representation by
repeated multiplication and the application of 
\begin{equation}
\chi _{a}(U)\chi _{b}(U)=\sum_{c}n\left( a,b;c\right) \chi _{c}(U) 
\end{equation}
where all $n$'s are non-negative integers. Taken together, these results
might suggest that all group characters are polynomials in $Tr_{F}U$ and its
complex conjugate. Consider, however, the product representation $N\otimes N$%
. It is reducible into $N(N+1)/2\oplus N(N-1)/2$, which are symmetric and
antisymmetric representations, respectively. Their characters are
respectively 
\begin{eqnarray}
\chi _{S} &=&\frac{1}{2}\left[ \left( Tr_{F}U\right) ^{2}+Tr_{F}\left(
U^{2}\right) \right] \nonumber\\
\chi _{A} &=&\frac{1}{2}\left[ \left( Tr_{F}U\right) ^{2}-Tr_{F}\left(
U^{2}\right) \right] \text{.}
\end{eqnarray}
Note that the sum $\chi _{S}+\chi _{A}$ is a polynomial in $Tr_{F}U$, but $%
\chi _{S}$ and $\chi _{A}$ are in general not. For example, in $SU(3)$ the
restriction on the eigenvalues of $U$ imposed by $\det \left( U\right) =1$
allows us to prove 
\begin{equation}
\frac{1}{2}\left[ \left( Tr_{F}U\right) ^{2}-Tr_{F}\left( U^{2}\right)
\right] =Tr_{F}\left( U^{\dagger}\right) 
\end{equation}
in accord with the $SU(3)$ result $3\otimes 3=6\oplus \overline{3}$. In this
case, it is true that $Tr_{F}\left( U^{2}\right) $ can be written as a
polynomial in $Tr_{F}U$ and $Tr_{F}U^{\dagger}$.
However, for $SU(4)$, unitarity
of $U$ gives instead 
\begin{equation}
\frac{1}{2}\left[ \left( Tr_{F}U\right) ^{2}-Tr_{F}\left( U^{2}\right)
\right] =\frac{1}{2}\left[ \left( Tr_{F}U\right) ^{2}-Tr_{F}\left(
U^{2}\right) \right] ^{\ast } 
\end{equation}
which shows that the $6$ representation of $SU(4)$ is real. $SU(N)$ $\,$%
group characters can be written as polynomials in $Tr_{F}\left( U\right) $
and its complex conjugate only for $SU(2)$ and $SU(3)$.

An alternative statement is that the Polyakov loop in the fundamental
representation, $Tr_{F}U$, is not sufficient to determine the eigenvalues of 
$U$, beginning with the case of $SU(4)$. Let us label the eigenvalues for
the Polyakov loop as $z_{i}$, with $i=1$ to $N$. For $SU(3)$, the
characteristic polynomial for the Polyakov loop is 
\begin{equation}
\prod_{i=1}^{3}(z-z_{i})=z^{3}-(z_{1}+z_{2}+z_{3})%
\,z^{2}+(z_{1}z_{2}+z_{2}z_{3}+z_{3}z_{1})\,z-z_{1}z_{2}z_{3}\text{.} 
\end{equation}
Since the determinant of a special unitary matrix is $1$, we have $%
z_{1}z_{2}z_{3}=1$, and the characteristic polynomial is 
\begin{equation}
z^{3}-z^{2}Tr_{F}U\,+zTr_{F}U^{+}-1 
\end{equation}
so for $SU(3)$, knowledge of $Tr_{F}U$ determines all eigenvalues, and the
free energy can be written as a function of $Tr_{F}U$ alone. For $SU(4)$,
similar considerations allow the characteristic polynomial to be written as 
\begin{equation}
z^{4}-z^{3}Tr_{F}U+z^{2}\frac{1}{2}\left[ \left( Tr_{F}U\right)
^{2}-Tr_{F}\left( U^{2}\right) \right] -zTr_{F}U^{+}+1 
\end{equation}
and knowledge of $Tr_{F}U$ must be supplemented by the value of $%
Tr_{F}\left( U^{2}\right) $. As $N$ increases, more information must be
supplied to reconstruct the eigenvalues.

\section{Model A}
\label{sec:modela}

We want to add terms to $f_{pert}$\vspace{1pt} which are subleading in
comparison to the $T^{4}$ behavior of $f_{pert}$, and introduce a mass scale
into $f$. This mass scale will determine the deconfinement temperature $%
T_{d} $. The first of our two models is obtained by adding, by hand, a mass
to the gauge bosons, and working with the high temperature expansion of the
resultant free energy.

As before, we have 
\begin{equation}
f=-\frac{1}{\beta }\sum_{j,k=1}^{N}2(1-\frac{1}{N}\delta _{jk})\int \frac{%
d^{3}k}{(2\pi )^{3}}\sum_{n=1}^{\infty }\frac{1}{n}e^{-n\beta 
\omega_{k}+in\Delta \theta _{jk}}\text{.} 
\end{equation}
but now $\omega _{k}=\sqrt{k^{2}+M^{2}}$. It is easy to derive the first two
terms in the high-temperature expansion. Higher order effects, which include
terms of order $TM^{3}$ and $M^{4}\ln \left( T/M\right) $, can be derived by
more sophisticated methods. Such terms were derived in the case of a trivial
Polyakov loop by Dolan and Jackiw \cite{Dolan:1974qd}.
See \cite{PeteMike:fttricks} for a relatively simple
derivation of the general case.\ The result is 
\begin{equation}
f_{A}\left( \theta \right) =-\sum_{j,k=1}^{N}\frac{1}{\pi ^{2}}(1-\frac{1}{N}%
\delta _{jk})\left[ -\frac{2\pi ^{4}}{3\beta ^{4}}B_{4}\left( \frac{\Delta
\theta _{jk}}{2\pi }\right) -\frac{M^{2}\pi ^{2}}{2\beta ^{2}}B_{2}\left( 
\frac{\Delta \theta _{jk}}{2\pi }\right) \right] 
\end{equation}
where the second Bernoulli polynomial $B_{2}(x)$ is given by $%
B_{2}(x)=x^{2}-x+1/6$ on the interval $\left( 0,1\right) $. The above
expression \textit{defines} the free energy for our model A.
In fact, the full one-loop free
energy for a massive gluon always favors $A_{0}=0$. 
We stress that
we view this derivation as merely providing
an indication of the type of additional terms that
might appear in the complete free energy. 
The Bernoulli
polynomials appear naturally in class functions which are almost
everywhere polynomials in the $\theta$'s; 
these class functions
are thus well-suited for the construction of a Landau
theory in the eigenvalues. In more explicit form, $f_{A}$ is given by

\begin{eqnarray}
f_{A}\left( \theta \right) &=&-T^{4}\sum_{j,k=1}^{N}(1-\frac{1}{N}\delta
_{jk})\left[ \frac{\pi ^{2}}{45}-\frac{1}{24\pi ^{2}}\left| \Delta \theta
_{jk}\right| _{2\pi }^{2}\left( \left| \Delta \theta _{jk}\right| _{2\pi
}-2\pi \right) ^{2}\right] \nonumber\\
&&+T^{2}M^{2}\sum_{j,k=1}^{N}(1-\frac{1}{N}\delta _{jk})\left[ \frac{1}{12}+%
\frac{1}{8\pi ^{2}}\left| \Delta \theta _{jk}\right| _{2\pi }\left( \left|
\Delta \theta _{jk}\right| _{2\pi }-2\pi \right) \right].
\end{eqnarray}
Because the free energy density $f_A$ is a class function of the Polyakov
loop eigenvalues by construction, $f_A$ is invariant under gauge transformations.
This important property will also hold for model B.

The $T^{4}$ term has a clear minimum when $\Delta \theta _{jk}=0$,
corresponding to $A_{0}=0$ and its $Z(N)$ equivalents, and will dominate for
large $T$. The $T^{2}$ term, however, will dominate for $T$ small and has a
global \textit{maximum} at $\Delta \theta _{jk}=0$. 
To explore the phase structure, it suffices to consider
real and positive $P_{F}$; it is convenient to use a parametrization for the 
$SU(N)$ angles which we will refer to as the $\phi $ parametrization. For $N$
even, we represent diagonal matrices as $diag\left[ \exp \left( i\phi
_{N/2},..,i\phi _{1},-i\phi _{1},..,-i\phi _{N/2}\right) \right] $, with the
eigenvalues ordered such that $\pi \geq \phi _{N/2}\geq ..\geq \phi _{1}\geq
0$. For $N$ odd, we represent diagonal matrices as $diag\left[ \exp \left(
i\phi _{\left( N-1\right) /2},..,i\phi _{1},0,-i\phi _{1},..,-i\phi _{\left(
N-1\right) /2}\right) \right] $, with the eigenvalues again ordered such
that $\pi \geq \phi _{\left( N-1\right) /2}\geq ..\geq \phi _{1}\geq 0$.

At low temperatures, the $\phi \,$will minimize 
\begin{equation}
2\sum_{j=2}^{N}\sum_{k=1}^{j}\,\left| \Delta \theta _{jk}\right| _{2\pi
}\left( \left| \Delta \theta _{jk}\right| _{2\pi }-2\pi \right) 
\end{equation}
For even $N$, this reduces to

\begin{equation}
\sum_{j=1}^{N/2}\left[ 4N\phi _{j}^{2}-8\pi \left( 2j-1\right) \phi
_{j}\right] 
\end{equation}
The minimum occurs at 
\begin{equation}
\phi _{j}=\frac{\pi \left( 2j-1\right) }{N} 
\end{equation}
which is precisely uniform spacing around the unit circle. For odd $N$, we
have a similar reduction to 
\begin{equation}
\sum_{j=1}^{\left( N-1\right) /2}\left[ 4N\phi _{j}^{2}-16\pi j\phi
_{j}\right] 
\end{equation}
and the minimum is given by 
\begin{equation}
\phi _{j}=\frac{2\pi j}{N}\text{.} 
\end{equation}
Note that in the case of $N\,$\ even, unitarity forces the eigenvalues away
from $\theta =0$, \textit{e.g.}, for $N=4$, the four angles form an ''X''
rather than a ''+''.

In this model, the deconfinement transition arises because of competition
between the $T^{4}$ term, which tends to force all eigenvalues to zero, and
the $T^{2}$ term, which forces the eigenvalues apart. We next discuss the
analytically tractable
cases of $SU(2)$ and $SU(3)$. For $SU(4)$ and higher, model A
is conveniently solved by numerical methods.

\subsection{Model A for $SU(2)$}

\vspace{1pt}In the case of $SU(2)$, we have only $\phi _{1}=\phi $ and $\phi
_{-1}=-\phi $. At low temperatures, we have $\phi =\pi /2$, giving $P_{F}=0$%
. The free energy is 
\begin{equation}
f_{A}=-\frac{\pi ^{2}T^{4}}{15}+\frac{4T^{4}}{3\pi ^{2}}\phi ^{2}\left( \phi
-\pi \right) ^{2}+\frac{M^{2}T^{2}}{4}+\frac{M^{2}T^{2}}{\pi ^{2}}\phi
\left( \phi -\pi \right) 
\end{equation}
This equation has an obvious symmetry under $\phi \leftrightarrow \pi -\phi $
associated with $Z(2)$ invariance. It is convenient to define a new variable 
$\psi =\pi /2-\phi $, which better manifests the $Z(2)$ symmetry. We obtain 
\begin{equation}
f=-\frac{\pi ^{2}T^{4}}{15}+\frac{T^{4}}{12\pi ^{2}}\left( \pi ^{2}-4\psi
^{2}\right) ^{2}+\frac{M^{2}T^{2}}{4}-\frac{M^{2}T^{2}}{4\pi ^{2}}\left( \pi
^{2}-4\psi ^{2}\right) 
\end{equation}
with $\psi =0$ representing confinement. The phase transition is second
order, in accord with the universality argument 
of Svetitsky and Yaffe \cite{Yaffe:1982qf}.
The deconfinement temperature $T_{d}$ is given by $%
T_{d}=\left( 3/2\right) ^{1/2}M/\pi \approx 0.38985M$, and $\psi \left(
T\right) $ behaves as 
\begin{equation}
\psi (T)=\sqrt{\frac{2T^{2}\pi ^{2}-3M^{2}}{8T^{2}}}
\end{equation}
Above $T_{d}$, the pressure $p=-f$ is given by 
\begin{equation}
p=\frac{1}{15}\pi ^{2}T^{4}-\frac{1}{4}M^{2}T^{2}+\frac{3}{16\pi ^{2}}M^{4}
\end{equation}
The internal energy is given by 
\begin{equation}
\varepsilon  =T\frac{dp}{dT}-p 
=\frac{16\pi ^{4}T^{4}-20M^{2}T^{2}\pi ^{2}-15M^{4}}{80\pi ^{2}}
\end{equation}
The dimensionless interaction measure $\Delta $,
defined from the stress-energy tensor as 
$ \Delta = (\langle T^\mu_\mu \rangle_T -\langle T^\mu_\mu \rangle_0)/T^4$, 
is given by 
\begin{equation}
\Delta =\frac{\varepsilon -3p}{T^{4}}
=\frac{M^{2}\left( T^{2}-T_{d}^{2}\right) }{2T^{4}}.
\end{equation}
Note for future reference that $\Delta $ falls off as $1/T^{2}$ 
above $T_{d}$.

This first, analytically tractable example exhibits the one real shortcoming
of these models: the pressure can go negative at low temperatures, and shows
non-monotonic behavior. A fully satisfactory theory would probably have the
pressure identically zero in the confined phase, or positive and very small
if glueball effects were included. We could add a small constant to the free
energy which would make $p(T_{d})=0$, at the cost of slightly changing the
graph of $p/T^{4}$ for temperatures just above $T_{d}$. 
Such a term is similar to the appearance of the bag constant in the bag model,
but with an important difference. In this model, any term independent of the 
temperature would
also be independent of the Polyakov loop eigenvalues, and would give the same
contribution in both the confined and deconfined phases.
The introduction of
this new parameter, moreover, does not necessarily 
render the pressure physical in the
region below $T_{d}$. Since our intent is to model the behavior above $T_{d}$%
, we view this behavior as a minor flaw.

\subsection{Model A for $SU(3)$}

\vspace{1pt}In the case of $SU(3)$, there are three eigenvalue $\phi
_{1}=\phi $, $0$ and $\phi _{-1}=-\phi $. At low temperatures, we have $\phi
=2\pi /3$, giving $P_{F}=0$. The free energy has the form 
\begin{eqnarray}
f_{A} &=&-T^{4}\frac{8\pi ^{2}}{45}+\frac{T^{4}}{6\pi ^{2}}\left[ 8\phi
^{2}\left( \phi -\pi \right) ^{2}+\phi ^{2}\left( \phi -2\pi \right)
^{2}\right] \nonumber\\
&&+\frac{2T^{2}M^{2}}{3}+\frac{T^{2}M^{2}}{2\pi ^{2}}\left[ 2\phi \left(
\phi -\pi \right) +\phi \left( \phi -2\pi \right) \right]
\end{eqnarray}
As in the case of $SU(2)$, a simple substitution is helpful. Defining $\psi
=2\pi /3-\phi $, we may eliminate the linear terms and write $f_{A}$ in the
form

\begin{equation}
f_{A}=\frac{8\pi ^{2}}{405}T^{4}+\left( \frac{3}{2\pi ^{2}}%
T^{2}M^{2}\allowbreak -\frac{2}{3}T^{4}\right) \psi ^{2}-\frac{2}{3\pi }%
T^{4}\psi ^{3}+\frac{3}{2\pi ^{2}}T^{4}\psi ^{4}\text{.} 
\end{equation}
The presence of a $\psi ^{3}$ term indicates that the phase transition will
be first order, as expected. The non-trivial minimum of $f_{A}$, which
represents the deconfined phase, is given by

\begin{equation}
\psi _{\min }=\frac{\pi T+3\sqrt{T^{2}\pi ^{2}-2M^{2}}}{6T} 
\end{equation}
The point at which $\psi _{\min }\,$develops an imaginary part is the
spinodal point, given by $T_{s}=\sqrt{2}M/\pi $. Below this temperature, the
deconfined phase is no longer metastable. There is another spinodal point,
associated with $\psi =0$. This temperature, which is given by $%
T_{s}^{\prime }=3M/2\pi $, is the temperature above which the confined phase
is no longer metastable.

A first order phase transition occurs when $f(\psi _{\min })=f(0)$, which
gives the critical value 
\begin{equation}
\psi =2\pi /9
\end{equation}
and this in turn implies the deconfinement temperature is 
\begin{equation}
T_{d}=\frac{9}{20\pi }\sqrt{10}M\approx \allowbreak .\,\allowbreak 452\,96M%
\text{.}
\end{equation}
Comparing the two spinodal temperatures to $T_{d}$, we have 
\begin{eqnarray*}
\frac{T_{s}}{T_{d}} &\approx &\allowbreak .\,\allowbreak 993\,81 \\
\frac{T_{s}^{\prime }}{T_{d}}\allowbreak  &\approx &1.\,\allowbreak 054\,1
\end{eqnarray*}
which indicates a very narrow range of metastability around $T_{d}$,
consistent with a very weak first order phase transition.

\vspace{1pt}

\section{\protect\vspace{1pt}Model B}
\label{sec:modelb}

\vspace{1pt}As will be demonstrated in section~\ref{sec:thermo},
model A gives a very
reasonable approximation to the free energy and associated thermodynamic
functions above $T_{d}$, as obtained from lattice simulations. It is
desirable to have a second model, to attempt to judge what features are
universal. Our second model, like the first, is obtained by physically
motivated considerations, but of a completely different type.

Let us suppose that there is some natural scale $R$ in position space over
which color neutrality is enforced. In other words, net color is allowed in
volumes of less than $R^{3}$, but the net color on larger scales is zero. We
think of space as being divided up into cells of size $R^{3}$, and assume
each cell is large enough that the conventional density of states may be
used. The general form of the partition function for a cell has been known
for some time
\cite{Redlich:1980bf,Turko:1981nr,Gorenstein:1983ib,Gorenstein:1983ua,Elze:1983du,Elze:1984un,Skagerstam:1983fq,Skagerstam:1984gv}
; it has the form 
\begin{equation}
Z_{cell}=\exp \left[ -\beta R^{3}f_{cell}\right] =\int \left( d\theta
\right) \exp \left[ -\beta R^{3}f_{pert}\left( \theta \right) \right] 
\end{equation}
where $\left( d\theta \right) $ denotes an integral over $SU(N)$ Haar
measure, and $f_{pert}(\theta )$ is as derived in section~\ref{sec:perturb}.
Although $f_{cell}$ 
turns out to have many of the properties associated with the
deconfinement transition, it lacks a true phase transition, as any model
based on integration over a finite number of variables must. In order to
have a phase transition, there must be cell-cell correlations. We construct
our model B by requiring that the eigenvalues $\theta $ be the same in all
cells. Such a condition might be derived by steepest descents, for example.
This leads to 
\begin{equation}
f_{B}\left( \theta \right) =f_{pert}\left( \theta \right) -\frac{1}{\beta
R^{3}}\ln \left[ J\left( \theta \right) \right] 
\end{equation}
where $J\left( \theta \right) $ is the Jacobian contained within the Haar
measure $\left( d\theta \right) $. For $SU(N)$, its explicit form is 
\begin{equation}
J\left( \theta \right) =\prod_{j<k}\sin ^{2}\left( \frac{\theta _{j}-\theta
_{k}}{2}\right) 
\end{equation}
up to a constant which is fixed by demanding that $\int \left( d\theta
\right) =1$. At temperatures large compared to $1/R$, minimization of $f_{B}$
over the $\theta _{j}$ variables gives a free energy close to $f_{cell}$.

At low temperatures, the free energy is dominated by the measure term, which
favors the same uniformly spaced pattern of eigenvalues found for model A.
This behavior is very familiar from random matrix theory
\cite{Itzykson:1989sy}.
At high temperatures, the free energy is dominated by $f_{pert}$%
, and a phase transition occurs because of the conflict between the two
terms. Physically, the conflict is between minimizing the energy of the
gauge boson gas and maximizing the entropy associated with color
fluctuations.

Model B clearly has features in common with model A. In both models, the
correct perturbative behavior is built in, and the deconfining phase
transition results from the interplay of two simple terms. Both models
introduce a single mass scale which determines $T_{d}$. However, the
physical motivation for the two models is rather different, and the mass
scale is introduced in different ways.

\section{$SU(N)$ \protect\vspace{1pt}Thermodynamics for $N=2,3,4$, and $5$%
}
\label{sec:thermo}

In this section, we present the pressure $p$, the energy density $%
\varepsilon $ and the interaction measure $\Delta \,$as functions of $T$ for 
$SU(2)$ through $SU(5)$ for both models. In the case of $SU(3)$, we also
compare the models with results from pure gauge simulations
\cite{Boyd:1996bx}.
As
discussed above, model A is analytically tractable in the cases of $SU(2)$
and $SU(3)$; for the case of $SU(4)\,$and higher, a numerical solution is
easily obtainable. Model B, because it involves transcendental functions,
must be solved numerically for all $N$.

For $SU(2)$, both models predict a second-order phase transition, in
agreement with universality arguments and lattice simulations
\cite{Svetitsky:1986ye}.
For $%
N=3,4$, and $5$, the phase transition is first order in both models.
Figures 1-3 show $p$, $\varepsilon $,
and $\Delta $ as a function of the dimensionless variable $T/T_{d}$ for the
case of $SU(3)$. Results from lattice simulations, model A, and model
B are shown. Since both models involve only a single dimensional parameter,
these graphs have no free parameters. While neither model is in precise
agreement with the simulation data, both are reasonable approximations over
the range $T_{d}-4.5\,T_{d}$. The most notable discrepancy is that both
models approach the blackbody limit faster than the simulation data.

It is enlightening to plot $p$, $\varepsilon $, and $\Delta $, each divided
by $N^{2}-1$, versus $T/T_{d}$ for $N=2$, $3$, $4$, and $5$. Figures 4-6
show the results for model A, and 7-9 for model B. Both models appear to be
quickly approaching a finite large-$N$ limit.

\vspace{1pt}Both the simulation data and the two models show power law
behavior in $\Delta $ for sufficiently large $T$, consistent with $\Delta
\propto 1/T^{2}$. This is the asymptotic behavior found analytically for $%
SU(2)$ in model A. For the sake of a consistent analysis, we fit the
simulation data and all models over the range $3T_{d}-4.5T_{d}$ to the
simple power law behavior $\Delta \propto 1/T^{p}$. The results are
presented in Table~\ref{table:1}.
The results are compatible with $1/T^{2}$ behavior
contaminated by subleading corrections. 
These results should be compared with the value $p=2.094(9)$ obtained
by us from an analysis of the simulation data of \cite{Boyd:1996bx}.
In fact, all of the curves can be
fit extremely well over the range $1.5T_{d}-4.5T_{d}$ by a curve of the form 
$\Delta =a/T^{2}+b/T^{4}$.

Considering that as formulated, models A and B have no free dimensionless
parameters, they do a very good job of representing thermodynamic behavior.
A more phenomenological approach would allow $M\,$and $R$, respectively, to
vary with temperature, allowing fits to lattice results.

\section{Quarks}
\label{sec:quarks}

\vspace{1pt}The effect of quarks can be straightforwardly included in an
approximate fashion by adding the free energy of quarks propagating in a
constant Polyakov loop background to the free energy of the gluons
\cite{Gross:1981br,Weiss:1981rj,Weiss:1982ev}.
It is very important, however, to note that this procedure neglects chiral
symmetry restoration. It is very likely that a unified model including
order parameters for both
deconfinement and chiral symmetry restoration is necessary to fully describe
the quark-gluon plasma\cite{Meisinger:1996kp,Dumitru:2001in}.
We limit
ourselves here to a discussion of very heavy quarks and the leading order
effect of light quarks. We will defer the subject of the interplay
of deconfinement and chiral symmetry restoration to a later time.

Quarks in the fundamental representation of the gauge group explicitly break 
$Z(N)$ symmetry. As the quark mass $m$ goes to infinity, this effect
vanishes. On the basis of theoretical models and lattice simulations, the
expected effect of very heavy quarks is to lower the critical temperature
\cite{Green:1984sd,Ogilvie:1984ss}.
This line of first-order critical points in the $T-m$
plane is expected to terminate in a second-order end point at some finite
value of $m$. A low-temperature expansion
\cite{Actor:1986zf,Actor:1987cf,PeteMike:fttricks}
gives the free energy of a massive quark as 
\begin{equation}
f_{q}\left( \theta \right) =\frac{2m^{2}T^{2}}{\pi ^{2}}\sum_{j=1}^{N}%
\sum_{n=1}^{\infty }\frac{\left( -1\right) ^{n}}{n^{2}}K_{2}\left( n\beta
m\right) \cos \left( n\theta _{j}\right) \text{,} 
\end{equation}
a form suitable for $\beta m$ sufficiently large.

We consider only the simple case of $SU(3)$ using model A with a single
heavy quark. Taking the parameter $M=1$, the deconfinement temperature in
the pure gauge theory is $T_{d}\left( m=\infty \right) =0.453M$. The
addition of a single heavy quark shifts the deconfinement temperature very
slightly, with the first-order line terminating at a quark mass of $m=2.57M$%
. At this critical end point, the deconfinement temperature is $T_{d}\left(
m=2.57M\right) =0.974\,T_{d}\left( m=\infty \right) =0.441\,M$.

For light quarks, the leading order behavior is proportional to $T^{4}$ and
independent of the quark mass, and therefore independent of chiral behavior.
It is given by \cite{Gross:1981br,Weiss:1981rj,Weiss:1982ev}
\begin{equation}
f_{q}\left( \psi \right) =N_{f}\left( -\frac{7\pi ^{2}}{4860}-\frac{20\pi }{%
81}\psi -\frac{1}{9}\psi ^{2}+\frac{4}{9\pi }\psi ^{3}-\frac{1}{6\pi ^{2}}%
\psi ^{4}\right) T^{4} 
\end{equation}
where $N_{f}$ denotes the number of light flavors. Recall that the confined
phase of the pure gauge theory was given by $\psi =0$. The term linear in $%
\psi \,$appearing in $f_{q}\left( \psi \right) \,$directly indicates the
breaking of the $Z(3)$ invariance of the pure gauge theory. Detailed
analysis shows that with $N_{f}=2$, the deconfinement transition is replaced
by a smooth crossover.
The only scale in this model is again the parameter $M$, which need not
have the same value found in the pure gauge theory, and a rapid
rise in all thermodynamic quantities begins at a temperature around $M/4$.

\section{Conclusions}
\label{sec:conclusion}

We have developed two phenomenological equations of state for the
quark-gluon plasma, which reproduce much of the thermodynamic behavior seen
in lattice simulations. These models have many attractive features. Not the
least of these is simplicity. Both models introduce a single new parameter.
For the most important case of $SU(3)$, the free energy is obtained
by minimizing over a single variable. 
Both models
correctly predict the order of the deconfining phase transitions for $SU(2)$
(second order) and $SU(3)$ (weakly first order). Both models predict first
order transitions in $SU(4)$ and $SU(5)$ as well. The pressure and other
thermodynamic quantities vary rapidly in the range $T_{d}-5T_{d}$ in both
models, and both appear to have smooth large $N$ limits.

Although the two models have very different phenomenological origins, the
numerical value of the parameters introduced are reasonable for the case of $%
SU(3)$. If we take the deconfinement temperature in a pure $SU(3)$ gauge
theory to be $270\,MeV$, then model A gives a value for $M$ of $596\,MeV$.
This is a plausible value for a constituent gluon mass. In model
B, we find that $R\,$is $1\,$fermi, which is of course a typical hadronic
scale. Thus in both models the phenomenological parameter introduced
has a reasonable value.

It is useful to compare these models to the naive Bag model equation of
state, which is in common usage in phenomenological applications
\cite{Cleymans:1986wb}.
The Bag
model also introduces a single dimensional parameter, the bag energy density 
$B$, but gives a first order transition for all $N$. The Bag model pressure
approaches the high-temperature, blackbody limit faster than both model A
and model B, and also lattice results. The graph of $\varepsilon /T^{4}$
versus $T\,$for the Bag model is monotonically decreasing, in complete
disagreement with lattice results. Finally, the Bag model predicts a $1/T^{4}
$ behavior for $\Delta $, which is ruled out by lattice results. As we have
seen, model A and model B both show a $1/T^{2}$ behavior in $\Delta $, which
is compatible with $SU(3)$ lattice data.

As we have formulated them, both model A and model B have no free parameters
once $T_{d}$ is fixed. It is clear that by allowing the parameters $M$ and $%
R $ to depend on the temperature, a better fit to lattice data can be
obtained at the cost of introducing additional phenomenological parameters.
We plan to explore this issue for the most important case of $SU(3)$.

The success of these phenomenological models strongly suggests a theoretical
point of view on the nature of the deconfinement mechanism, independent of
the specific gauge theory under study. In confining theories, such as pure
gauge theories, the eigenvalue distributions of the Polyakov loop are peaked
at low temperature around values evenly spaced about the unit circle in
such a way that the expectation value of the Polyakov loop is zero. Such
behavior is characteristic of random $SU(N)\,$matrices. At the deconfining
transition temperature, the peaks of the eigenvalue distributions change,
and move towards $1$, which is the asymptotic limit as $T$ goes to infinity.
The consistency of this picture can be tested using lattice Polyakov loop
eigenvalue distributions for $SU(4)$ and higher. In theories with light
quarks, the deconfinement transition may be replaced by a rapid crossover,
which is again associated with the motion of the Polyakov loop eigenvalues.
We believe that this picture of deconfinement coupled with a field-theory
inspired model of chiral symmetry breaking has the potential to fully model
the equation of state of the quark-gluon plasma.

Associating deconfinement with changes in Polyakov loop eigenvalue
distributions gives some perspective on recent attempts to model the plasma
equation of state using the hard thermal loop (HTL)\ approximation
\cite{Andersen:2000sf,Andersen:2000va,Blaizot:1999ap,Blaizot:2001fc}. 
The HTL
approximation is a resummation of high-temperature perturbation theory, and
may provide a good explanation of the slow asymptotic approach to the
blackbody limit at very high temperatures. However, if the picture we have
advocated here is correct, then no approach based on perturbation theory can
explain the behavior of the plasma in the crucial region from $T_{d}$ to $%
5T_{d}$, because the Polyakov loop eigenvalues are assumed to be $1$. In
fact, preliminary numerical simulations of pure $SU(3)$ gauge theories using 
$N_{t}=4$ lattices indicate that the eigenvalues of individual Polyakov
loops move much more slowly towards $1\,$than either model predicts. It is
possible that a hybrid approach combining hard thermal loops with the
minimal amount of phenomenology used in our models would be quite successful
in reproducing the quark-gluon plasma thermodynamics over the entire range
of temperatures above $T_{d}$.

\vspace{1pt}Finally, there is the question of the connection of these models
to candidate explanations for confinement having a deeper basis in the
underlying gauge theory dynamics. Both model A and model B attribute
confinement to a single, simple term in the free energy. Derivation
of a similar term from one of these more fundamental candidate explanations
of confinement would be very satisfying. A proper field-theoretic basis for
the motion of the Polyakov loop eigenvalues would answer many questions.

\begin{figure}
\includegraphics[width=5in]{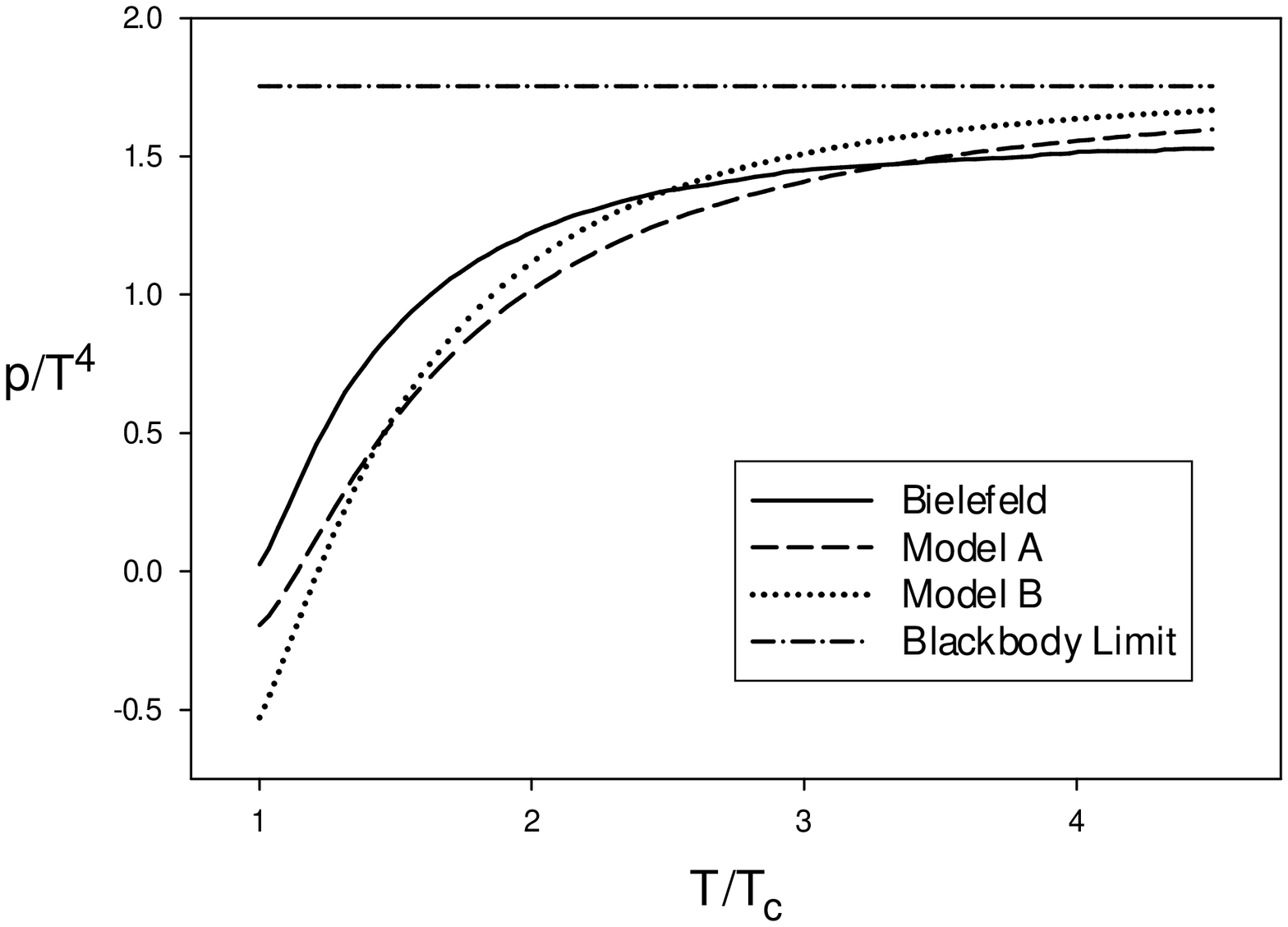}
\caption{$p/T^4$ versus $T/T_c$ for $SU(3)$}
\label{su3-p}
\end{figure}
\begin{figure}
\includegraphics[width=5in]{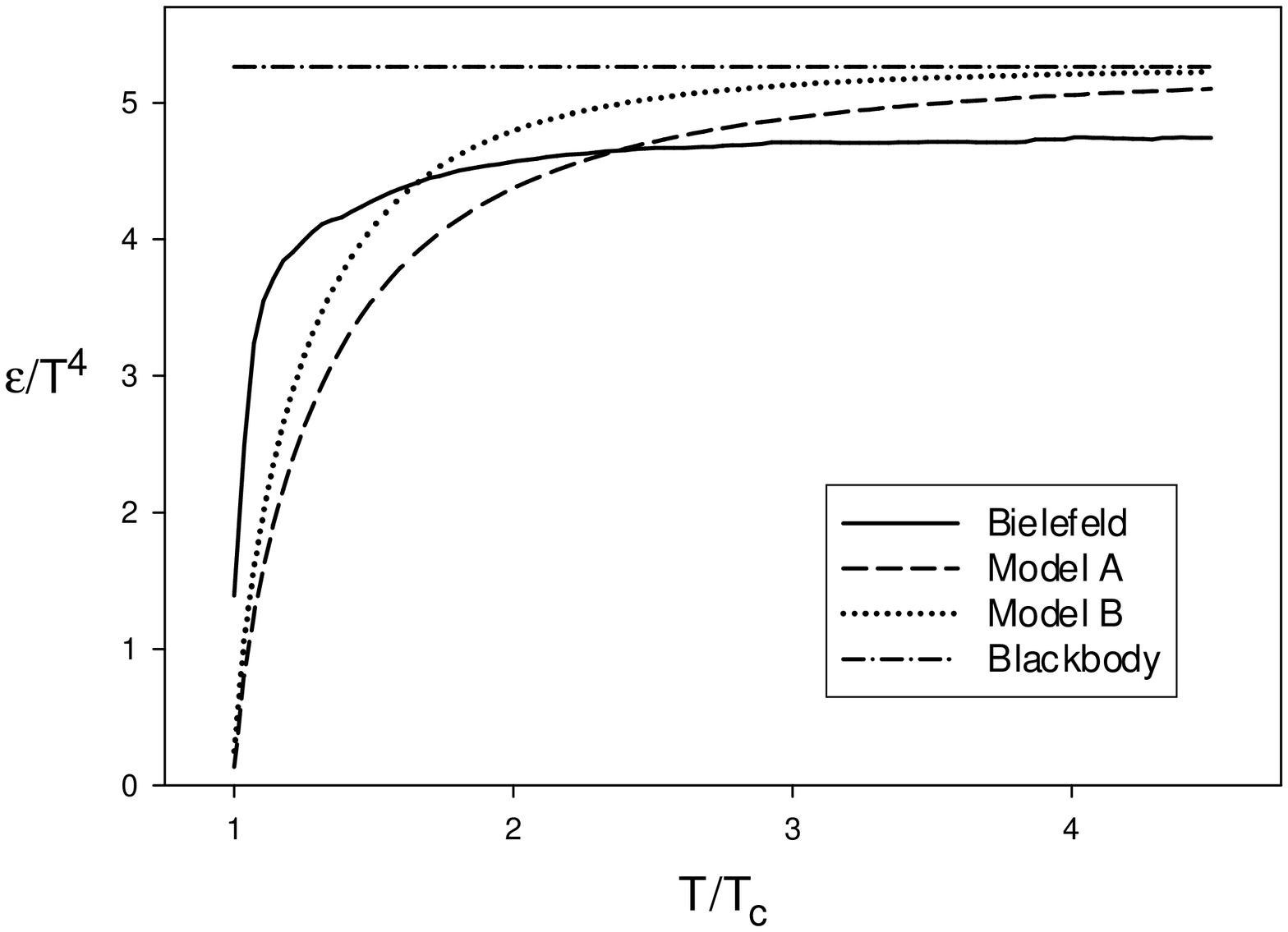}
\caption{$e/T^4$ versus $T/T_c$ for $SU(3)$}
\label{su3-e}
\end{figure}
\begin{figure}
\includegraphics[width=5in]{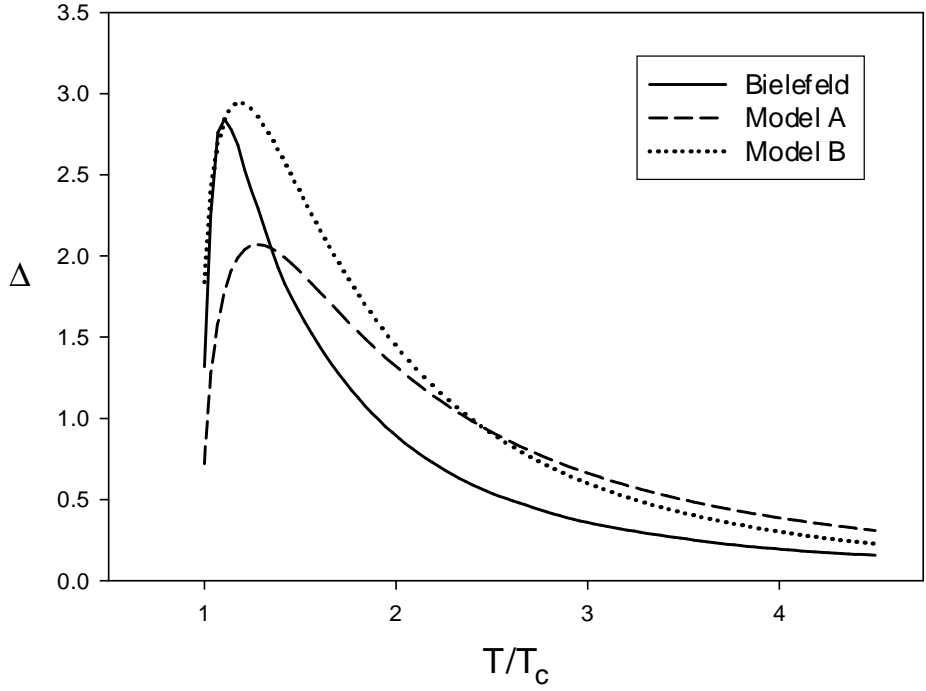}
\caption{$\Delta/T^4$ versus $T/T_c$ for $SU(3)$}
\label{su3-d}
\end{figure}

\begin{figure}
\includegraphics[width=5in]{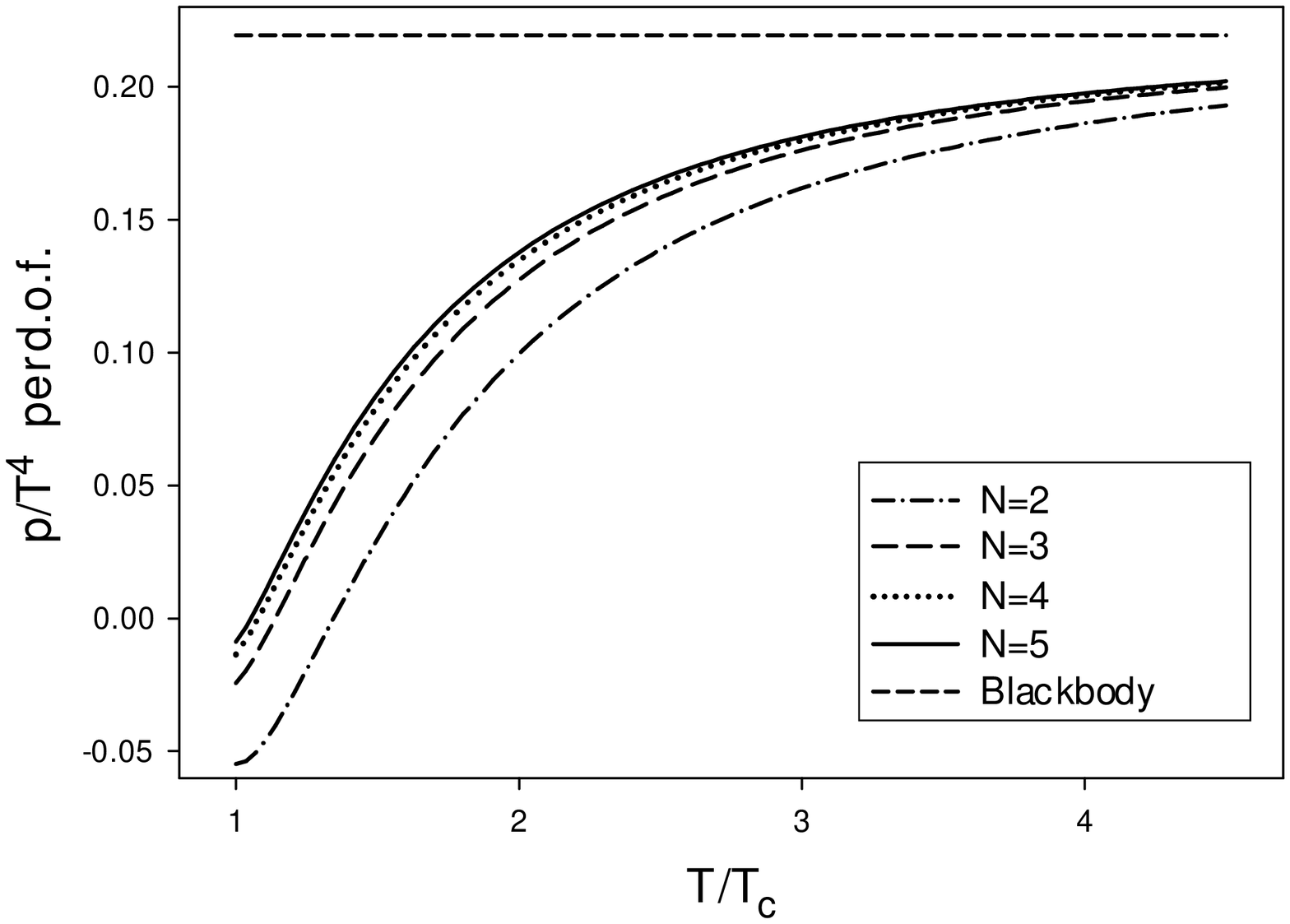}
\caption{$p/T^4$ versus $T/T_c$ using model A for $N=2,3,4,5$}
\label{b2-p}
\end{figure}
\begin{figure}
\includegraphics[width=5in]{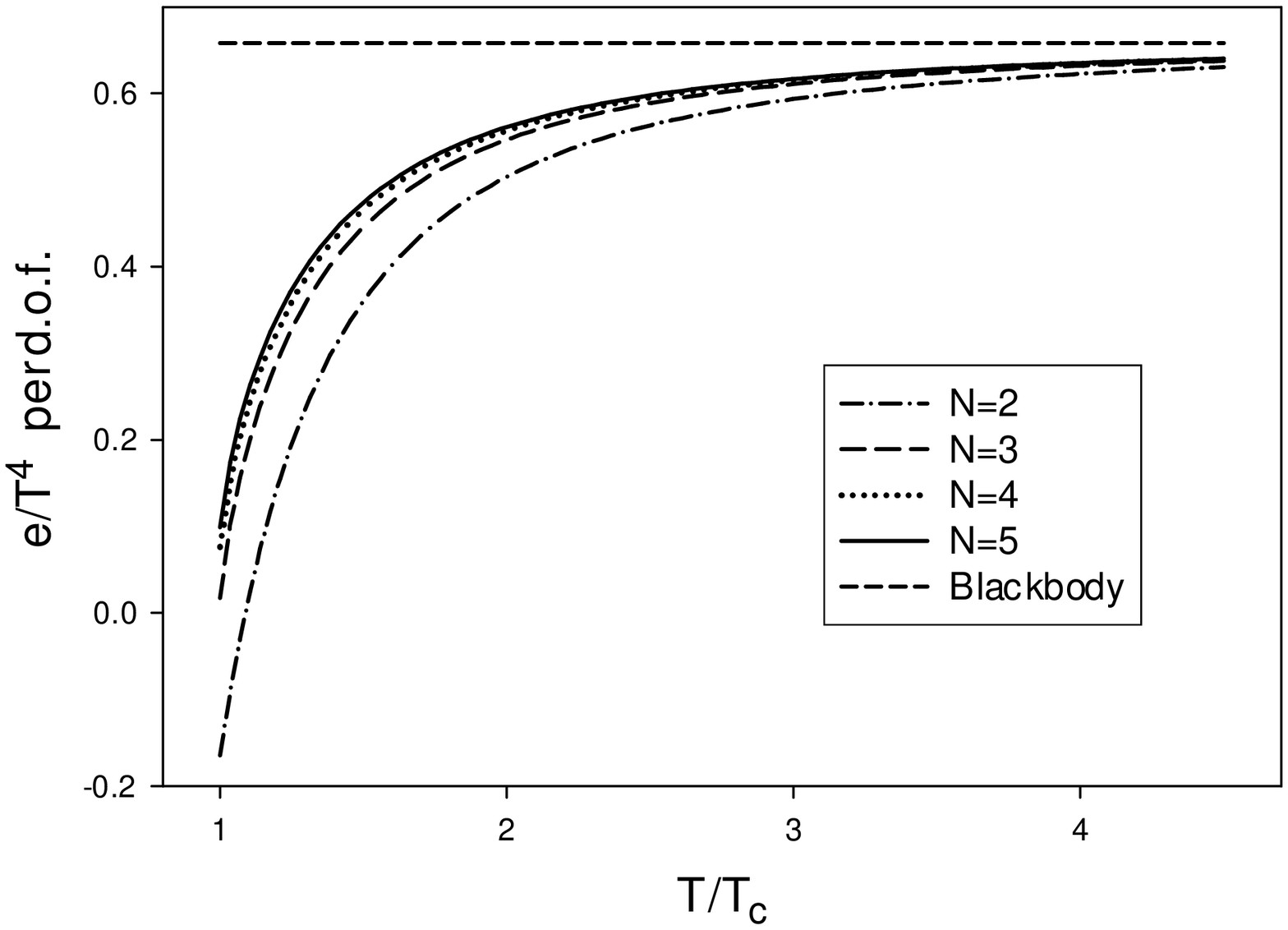}
\caption{$e/T^4$ versus $T/T_c$ using model A for $N=2,3,4,5$}
\label{b2-e}
\end{figure}
\begin{figure}
\includegraphics[width=5in]{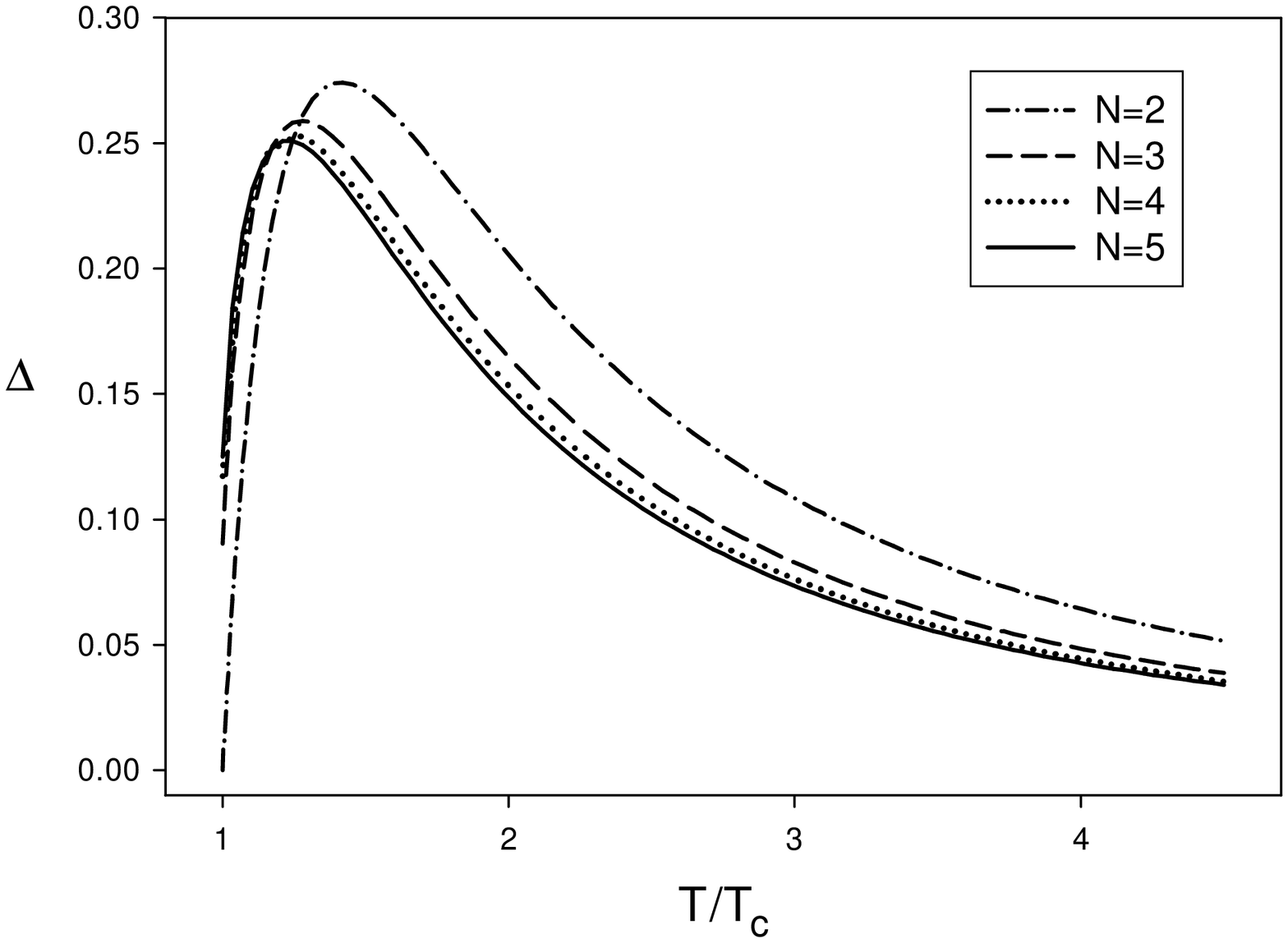}
\caption{$\Delta/T^4$ versus $T/T_c$ using model A for $N=2,3,4,5$}
\label{b2-d}
\end{figure}  

\begin{figure}
\includegraphics[width=5in]{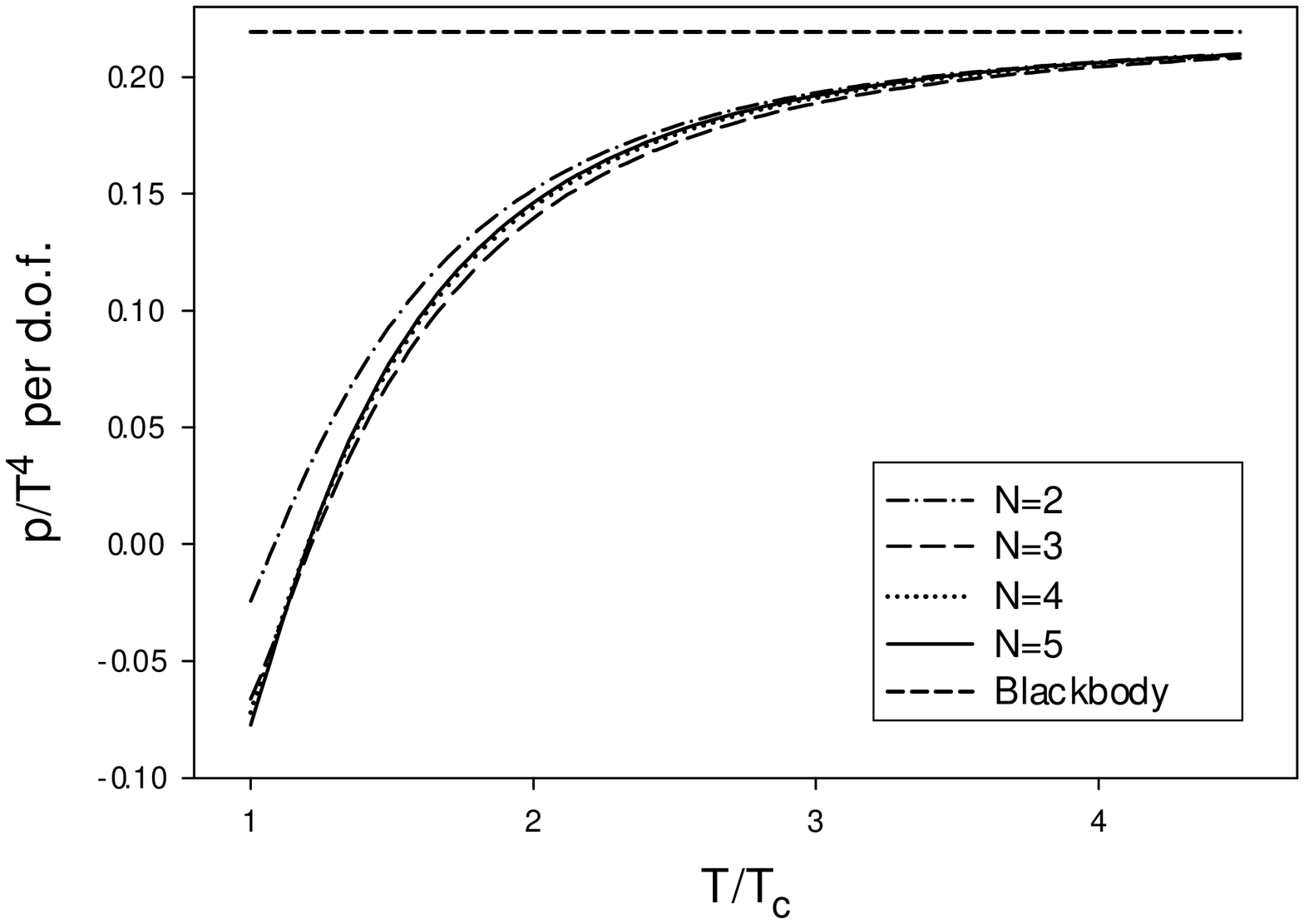}
\caption{$p/T^4$ versus $T/T_c$ using model B for $N=2,3,4,5$}
\label{meas-p}
\end{figure}
\begin{figure}
\includegraphics[width=5in]{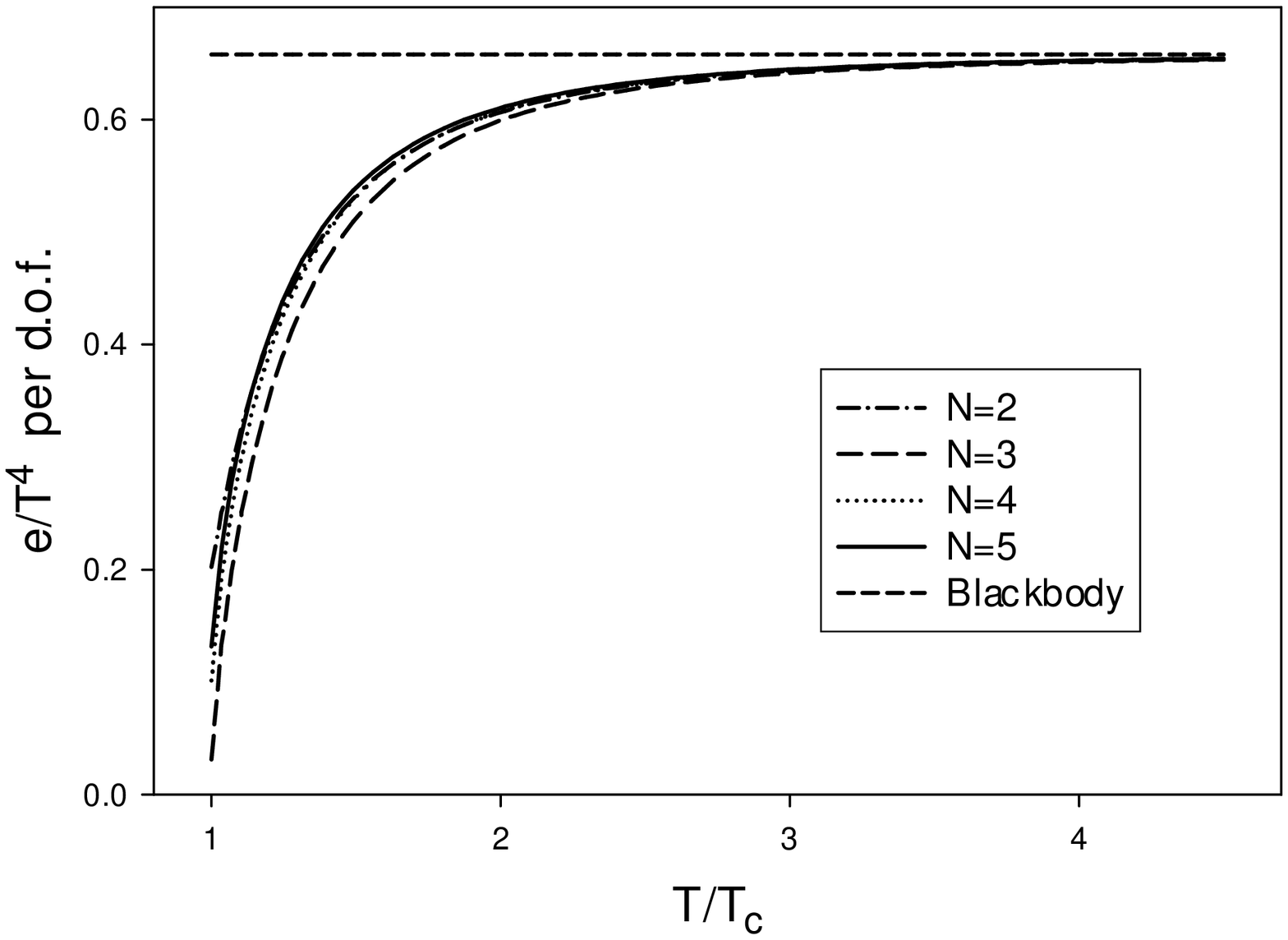}
\caption{$e/T^4$ versus $T/T_c$ using model B for $N=2,3,4,5$}
\label{meas-e}
\end{figure}
\begin{figure}
\includegraphics[width=5in]{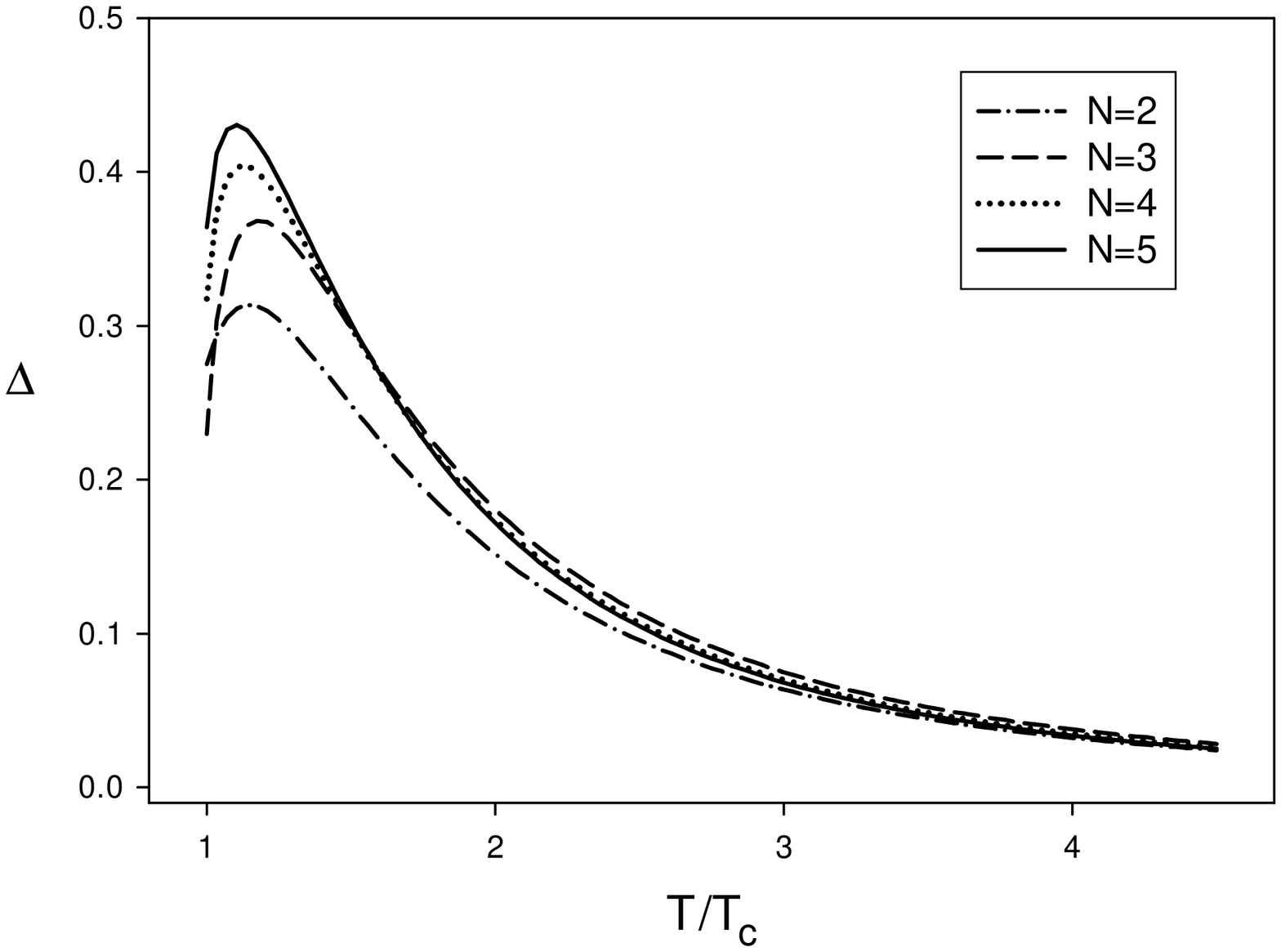}
\caption{$\Delta/T^4$ versus $T/T_c$ using model B for $N=2,3,4,5$}
\label{meas-d}
\end{figure} 

\begin{table}
\caption{The parameter $p$ obtained from the fit $\Delta \propto 1/T^{p}$.}
\label{table:1}
\begin{tabular}{ll}
Model A N=2 & 1.833(3) \\ 
Model A N=3 & 1.878(2) \\ 
Model A N=4 & 1.889(2) \\ 
Model A N=5 & 1.893(2) \\ 
Model B N=2 & 2.377(4) \\ 
Model B N=3 & 2.398(3) \\ 
Model B N=4 & 2.436(3) \\ 
Model B N=5 & 2.457(3)
\end{tabular}
\end{table}

\end{document}